\pdfoutput=1
\documentclass[sigplan,screen,nonacm,twocolumn]{acmart}
\RequirePackage{etoolbox}

\providebool{fullversion}
\providebool{arxivversion}

\usepackage{preamble}
\usepackage{iris}
\usepackage{prob}
\usepackage{examples}

\usepackage{listings}

\newcommand\leaninline{\lstinline[language=lean]}
\newcommand\rocqinline{\lstinline[language=coq]}

\usepackage{color}
\definecolor{keywordcolor}{rgb}{0.7, 0.1, 0.1}   \definecolor{tacticcolor}{rgb}{0.0, 0.1, 0.6}    \definecolor{commentcolor}{rgb}{0.4, 0.4, 0.4}   \definecolor{symbolcolor}{rgb}{0.0, 0.1, 0.6}    \definecolor{sortcolor}{rgb}{0.1, 0.5, 0.1}      \definecolor{attributecolor}{rgb}{0.7, 0.1, 0.1} 

\ifbool{fullversion}{
\newcommand{\appref}[1]{\cref{#1}}
  \newcommand{\Appref}[1]{\Cref{#1}}
}{
\newcommand{\appref}[1]{the Appendix}
  \newcommand{\Appref}[1]{The Appendix}
}

\ifbool{arxivversion}{
\usepackage{caption}
}{
\usepackage[belowskip=-10pt,aboveskip=4pt]{caption}
\setlength{\intextsep}{15pt} }

\makeatletter
\ifbool{arxivversion}{}{
\let \MathparLineskip \mpr@lesslineskip }
\makeatother

\ifbool{arxivversion}{
  \newcommand{\vsquish}[1]{}
}{
  \newcommand{\vsquish}[1]{\vspace{-#1}}
}

\usepackage{tcolorbox}
\tcbuselibrary{skins,breakable}
  \newtcolorbox{result}{
    blanker,
    extras={interior engine=spartan},
    grow to left by=2pt,left*=0mm,
    grow to right by=2pt,right*=0mm,
    top=1mm,bottom=1mm,
    beforeafter skip balanced=0.1\baselineskip plus 2pt,
    breakable,
    colback=cyan!40
  }

\usepackage{pgfplots}
\pgfplotsset{compat=1.18}

\ifbool{fullversion}{
}{}

\begin{document}

\title{Iris in Lean}

\author[M. de Medeiros]{Markus de Medeiros}
\orcid{0009-0005-3285-5032}
\affiliation{\institution{New York University}
  \city{New York}
  \country{USA}
}
\email{mjd9606@nyu.edu}

\author[S. Stepanenko]{Sergei Stepanenko}
\orcid{0000-0002-7322-5644}
\affiliation{\institution{Aarhus University}
  \city{Aarhus}
  \country{Denmark}
}
\email{sergei.stepanenko@cs.au.dk}

\author[Z. Liu]{Zongyuan Liu}
\orcid{0000-0001-9652-4869}
\affiliation{\institution{Aarhus University}
  \city{Aarhus}
  \country{Denmark}
}
\email{zy.liu@cs.au.dk}

\author[O. Soeser]{Oliver Soeser}
\orcid{0009-0006-8475-3676}
\affiliation{\institution{ETH Zurich}
  \city{Zurich}
  \country{CH}
}
\email{osoeser@ethz.ch}

\author[F. Leal]{Fernando Leal}
\orcid{0009-0005-1282-1089}
\affiliation{\institution{ENS Paris-Saclay}
  \city{Paris}
  \country{FR}
}
\email{fernando.leal_sanchez@ens-paris-saclay.fr}

\author[A. Tang]{Alvin Tang}
\orcid{0009-0001-1041-5360}
\affiliation{\institution{Institute of Science and Technology Austria (ISTA)}
  \city{Klosterneuburg}
  \country{Austria}
}
\email{alvin.tang@ista.ac.at}

\author[M. Vistrup]{Max Vistrup}
\orcid{0009-0000-6739-4890}
\affiliation{\institution{ETH Zurich}
\city{Zurich}
  \country{CH}
}
\email{max.vistrup@inf.ethz.ch}

\author[R. Jung]{Ralf Jung}
\orcid{0000-0001-7669-6348}
\affiliation{\institution{ETH Zurich}
\city{Zurich}
  \country{CH}
}
\email{ralf.jung@inf.ethz.ch}

\author[M. Carneiro]{Mario Carneiro}
\orcid{0000-0002-0470-5249}
\affiliation{\institution{Chalmers}
  \city{Gothenburg}
  \country{Sweden}
}
\email{marioc@chalmers.se}

\author[J. Tassarotti]{Joseph Tassarotti}
\orcid{0000-0001-5692-3347}
\affiliation{\institution{New York University}
  \city{New York}
  \country{USA}
}
\email{jt4767@nyu.edu}

\author[M. Sammler]{Michael Sammler}
\orcid{0000-0003-4591-743X}
\affiliation{\institution{Institute of Science and Technology Austria (ISTA)}
  \city{Klosterneuburg}
  \country{Austria}
}
\email{michael.sammler@ista.ac.at}

\author[L. Birkedal]{Lars Birkedal}
\orcid{0000-0003-1320-0098}
\affiliation{\institution{Aarhus University}
  \city{Aarhus}
  \country{Denmark}
}
\email{birkedal@cs.au.dk}

\begin{abstract}

The Iris framework for concurrent separation logic has been widely used for program verification research.
An important factor contributing to the framework's adoption is its high-quality mechanization in Rocq.
This mechanization uses a number of Rocq features in sophisticated ways, including a carefully constructed algebraic hierarchy for modeling separation logic resources, and a proof mode for embedded separation logic proofs, which combines custom Ltac with extensible typeclasses.
The library has been developed for over a decade with dozens of contributors, with an emphasis on modularity and maintainability.

This paper presents Iris-Lean, a reimplementation of Iris in the Lean proof assistant.
The reimplementation provides complete coverage of all of the non-experimental features included in Iris.
We explore how Lean features like flexible metaprogramming and quotient types can simplify the design and usage of Iris.
Using these features, we provide a novel implementation of the Iris proof mode with improved performance, simplify the handling of equivalences through quotient types,
build a variant of Diaframe proof automation, and provide convenience features like automatic construction of Iris fixed points.
Building on Lean lets us integrate with the extensive Mathlib library, allowing us to re-use results from this library for program verification tasks that have heavy mathematical dependencies, as we demonstrate with an application to probabilistic program verification.

\end{abstract}

\maketitle

\section{Introduction}
\label{sec:introduction}

Iris~\citep{iris,iris2,iris3,irisjournal} is a framework for higher-order
concurrent separation logic implemented and verified in Rocq. Iris has been used
to define and implement separation logics for a wide variety of programming
languages; see, \eg \citep{10.1145/3158154,
10.1007/978-3-030-17184-1_3,
10.1145/3371074,
10.1007/978-3-030-44914-8_13,
DBLP:journals/pacmpl/GiarrussoSTBK20,
DBLP:journals/pacmpl/MevelJP20,
DBLP:journals/jacm/GeorgesGSTTDB24,
273703,
10.1145/3519939.3523434,
DBLP:journals/pacmpl/RaoGLWPGB23,
DBLP:journals/pacmpl/VilhenaP21,
DBLP:journals/pacmpl/0001HMLGTB24,
DBLP:journals/pacmpl/Seassau0MP25}, and to develop semantic models of
sophisticated type systems; see, \eg \citep{jacm-iris-logrel,DBLP:journals/lmcs/FruminKB21,
DBLP:journals/pacmpl/GregersenBTB21,
DBLP:journals/pacmpl/TimanySKB18,
10.1145/3656422,
DBLP:journals/pacmpl/GregersenAHTB24,
DBLP:conf/esop/VilhenaP23}. While Iris
has mostly been used to reason about safety properties, several projects have
also investigated applications to reason about security properties such as
non-interference, as well as probabilistic properties and liveness properties; see
\href{https://iris-project.org/}{iris-project.org} for a complete overview. In
all of these projects, the Rocq formalization of Iris has been crucial, as it has
made it much easier to formalize new Iris-related projects. Notably, this
formalization not only covers the meta-theory (\eg
soundness of a program logic) but also provides the \emph{Iris Proof Mode} (IPM)~\citep{DBLP:conf/popl/KrebbersTB17,mosel} to facilitate
reasoning inside the embedded logic using tactics, much like one reasons in
ordinary Rocq. The IPM is extensible, which is used by many of the aforementioned projects
to implement support for their own domain-specific logical connectives.

In this paper we present the Iris-Lean project, a reimplementation of Iris in
the Lean proof assistant. Given the mentioned success of Iris in Rocq, one may
wonder about the motivation for such a reimplementation.
In fact, there are several motivating factors:
(1) While the Iris Proof Mode is a feat of engineering, implemented using
tricky Ltac code, Lean offers outstanding facilities for metaprogramming, which may make it
easier to implement and extend proof modes, and to explore different design decisions.
(2) Lean may support more automation, either by using built-in tactics, such as \leaninline|grind|,
or by programming automated solvers using Lean's metaprogramming features.
(3) Lean has a very extensive library, Mathlib~\citep{mathlib,DBLP:conf/cpp/X20}, with nearly 2,500,000 lines of formalized mathematics, much of which has yet to be ported to the Rocq Prover.
Access to this library can help us explore program logics which make use of more sophisticated mathematical techniques.
(4) By implementing Iris in a new proof assistant, one gets to
re-evaluate different implementation decisions and possibly explore some new ones.
(5) Finally, one might hope that Lean's focus on performance
makes an implementation of Iris in Lean more performant than the existing Rocq
implementation of Iris.

We will begin with an overview (\cref{sec:overview}) of the
reimplementation of Iris in Lean, which covers all of the Iris logic and Proof Mode (excluding generalized step indices).
To evaluate using Iris-Lean on a full-fledged program logic, we have also ported more than 80\% of the separation logic for the HeapLang programming language, including a variety of examples.
After the overview, we describe some
of the technical differences between the Iris-Lean and the Iris-Rocq
implementations (\cref{sec:mechanization}), and then we move on
to describing some of the new applications of Iris-Lean (\cref{sec:evaluation}),
including a new probabilistic program logic, automation for proof search and for declaring fixed points.
We also describe some performance benefits. Throughout these sections, we will
return to the list of motivations above.
Finally, we will conclude and discuss related work (\cref{sec:conclusion}).

 \section{Overview}
\label{sec:overview}

\begin{figure*}[t]
\captionsetup[subfigure]{belowskip=6pt,aboveskip=2pt}
\setlength{\abovecaptionskip}{0pt}
\begin{minipage}[t]{0.45\textwidth}
\begin{lstlisting}[language=lean,aboveskip=0pt,belowskip=0pt, basicstyle=\footnotesize\ttfamily,
    numbers=left,numberstyle=\tiny\color{commentcolor},numbersep=4pt,
    xleftmargin=1.4em,escapeinside={(*}{*)}]
def newcounter := hl_val^ λ _, ref(#0)(*\label[line]{line:scope}*)
def incr := hl_val^
  rec incr l :=(*\label[line]{line:binder}*)
    let n := !l;
    if snd(cmpXchg(l, n, #1 + n))
      then #() else incr l
def read := hl_val^ λ l, !l

theorem newcounter_mono_spec :
  ⊢@{IProp GF} {{ True }} hl(&newcounter #())(*\label[line]{line:entail}*)
    {{ l, RET hl_val(#l); mcounter N l 0 }} :=
by
  unfold newcounter; iintro ^Φ - HΦ(*\label[line]{line:unfold}*)
  wp_lam; wp_alloc l with Hl(*\label[line]{line:alloc}*)
  imod (MonoNat.own_alloc 0) with ⟨^γ, Hγ, Hγ'⟩(*\label[line]{line:with}*)

  imod (inv_alloc N _ (mcounter_inv γ l)) $$ [Hl Hγ]
  · inext; iexists 0; iframe
  imodintro; iapply HΦ; unfold mcounter; iframe(*\label[line]{line:unfoldlast}*)
(*\strut*)
\end{lstlisting}
\vfill
\subcaption{Iris-Lean}
\label{fig:monotone-counter:lean}
\end{minipage}\hfill
\begin{minipage}[t]{0.55\textwidth}
\begin{lstlisting}[language=coq,frame=none,aboveskip=0pt,belowskip=0pt,literate=
    {λ}{{\color{symbolcolor}\(\lambda\)\color{black}}}1
    {Φ}{{\(\Phi\)}}1
    {γ}{{\(\gamma\)}}1
    {●}{{\(\bullet\)}}1
    {◯}{{\(\circ\)}}1
    {⋅}{{\(\cdot\)}}1,
    basicstyle=\footnotesize\ttfamily]
Definition newcounter : val := λ: <>, ref #0.
Definition incr : val :=
  rec: "incr" "l" :=
    let: "n" := !"l" in
    if: Snd (CmpXchg "l" "n" (#1 + "n"))
      then #() else "incr" "l".
Definition read : val := λ: "l", !"l".

Lemma newcounter_mono_spec :
  {{{ True }}} newcounter #()
    {{{ l, RET #l; mcounter l 0 }}}.
Proof.
  iIntros (Φ) "_ HΦ". rewrite /newcounter /=.
  wp_lam. wp_alloc l as "Hl".
  iMod (own_alloc (● (MaxNat O) ⋅ ◯ (MaxNat O))) as (γ) "[Hγ Hγ']";
    first by apply auth_both_valid_discrete.
  iMod (inv_alloc N _ (mcounter_inv γ l) with "[Hl Hγ]").
  { iNext. iExists 0. by iFrame. }
  iModIntro. iApply "HΦ". rewrite /mcounter; eauto 10.
Qed.
\end{lstlisting}
\subcaption{Iris-\rocq}
\label{fig:monotone-counter:rocq}
\end{minipage}
\caption{A HeapLang monotone counter, specification, and proof in Iris-Lean (left) and Iris-\rocq (right).}
\Description{Side-by-side code listings of the same monotone-counter definition, specification and proof, in Iris-Lean on the left and Iris-Rocq on the right.}
\label{fig:monotone-counter}
\end{figure*}

Iris-Lean is a faithful port of Iris to Lean~4. While Iris-Lean is written from scratch, we track its correspondence to Iris-\rocq on a per-declaration basis (detailed in \cref{sec:status:coverage}), including our departures, which we elaborate on in~\cref{sec:mechanization}.
Finally, \cref{sec:heaplang} illustrates some key features of Iris-Lean with an end-to-end example from HeapLang, taking a program from its syntax to a closed statement about its operational semantics.

Iris-Lean comprises the Iris library of ghost resources, the step-indexed bunched implication (BI) interface \citep{ohearn99bi}, the Iris base logic, the Iris Proof Mode (IPM), the library of derived \leaninline|IProp| constructions, the Iris language interface and the language-parametric program logic, and the HeapLang example program logic.
The port of the core Iris library (\ie not HeapLang) is complete,\footnote{Relative to Iris-\rocq revision \texttt{a5375188} of August 2026.} in that every Iris-\rocq definition either has a Lean counterpart or is deliberately handled differently in Lean.
The only exception is generalized step indices, for which we are still investigating how to best balance concision, performance and robustness.

To illustrate Iris-Lean in action, \Cref{fig:monotone-counter} shows the higher-order specification and proof of a simple concurrent counter in HeapLang, realized both in Iris-Lean and Iris-Rocq.
The Iris-Lean tactics closely resemble their Iris-\rocq counterparts, though with some minor Lean-specific adaptations.
Most visibly, no strings occur in the surface syntax of Iris-Lean where Iris-Rocq uses them, as binders (\lineref{line:binder}) and hypothesis names (\lineref{line:alloc}) are ordinary Lean identifiers. Some parts are more verbose: HeapLang syntax is entered through the explicit macros \leaninline|hl_val| and \leaninline|hl| for notation scoping (\lineref{line:scope}), since Lean cannot associate a scope with a type and open it automatically. The entailment further carries an explicit type ascription \leaninline|@{IProp GF}| (\lineref{line:entail}).
Nevertheless, the proofs written by users of Iris-Lean remain in intentionally close correspondence with their Iris-Rocq counterparts.

\subsection{The Mechanics of Porting Iris-\rocq}
\label{sec:status:coverage}
Early on in the port, we discovered that it was useful to systematically track the correspondence between Iris-\rocq and Iris-Lean at the level of individual declarations.
To facilitate this we developed a custom attribute, \leaninline|rocq_alias|, which can be attached to any definition, theorem, or structure field in Iris-Lean.
Tagging any declaration with \leaninline|@[rocq_alias Name]| automatically registers it as a new alias under \leaninline|Rocq.Name|, which is given an informative \emph{deprecation message} guiding any user of \leaninline|Rocq.Name| to its Lean counterpart.
We used this feature extensively during the port, but even with the port complete, it remains useful for clients of Iris-Lean porting other Iris-based program logics.

This explicit correspondence also enables fine-grained tracking of our porting progress.
On each commit to the \lstinline|master| branch of the Iris-Lean repository, the CI (1) compares all known aliases from Lean against a pinned version of Iris-\rocq{}, (2) updates a website summarizing the correspondence and overall progress, and (3) generates a GitHub issue if our aliases have gone out of date.
We view \lstinline|rocq_alias| as an important tool we can use to keep Iris-Lean in sync with Iris-\rocq{} in the future.

\subsection{HeapLang}
\label{sec:heaplang}

\begin{figure}
\begin{lstlisting}[language=lean,aboveskip=0pt,belowskip=0pt, basicstyle=\footnotesize\ttfamily]
def quicksort : Val := hl_val^
  rec quicksort l :=
    match l with
    | none() => l
    | some(x) =>
      let p := !x;
      let part := &partition (fst(p)) (snd(p));
      let a := quicksort (fst(part));
      let b := quicksort (snd(part));
      &append a (&cons (fst(p)) b)

theorem quicksort_spec l ls :
    {{ isList (GF := GF) l ls }}
      hl(&quicksort &l)
    {{ l' ls', RET l'; isList l' ls' ∗
      ⌜Pairwise LE.le ls'⌝ ∗ ⌜ls ~ ls'⌝ }} := by
  iloeb as IH generalizing ^l ^ls
  iintro ^Φ Hl HΦ; wp_rec; rw [isList.eq_def]
  cases ls with dsimp only
  | nil =>
    icases Hl with ^rfl; wp_pures; imodintro
    iapply HΦ $$ ^_ ^([]) <;> simp [isList] <;> itrivial
  | cons head tail =>
    icases Hl with ⟨^l, ^tl, ^rfl, Hpt, Hl⟩
    wp_load
    wp_smart_apply partition_spec $$ [$] with ^l1 ^l2 H1 H2
    wp_smart_apply IH $$ [$H1] with ^l1' ^ls1' ⟨H1, ^_, ^_⟩
    wp_smart_apply IH $$ [$H2] with ^l2' ^ls2' ⟨H2, ^_, ^_⟩
    wp_smart_apply cons_spec $$ H2 with ^_ Hcons
    wp_smart_apply append_spec $$ [$H1 Hcons] with ^_ _
    iapply HΦ; iframe; isplit <;> ipureintro
    · have : ls2'.all (head < ·) := by grind
      grind [pairwise_cons]
    · grind [filter_append_perm]

def sortAndCheck (l : List Int) : Exp := hl^
  let v := &(makeList l); let v' := &quicksort v;
  &checkSorted (none()) v'

theorem sortAndCheck_spec (l : List Int) :
    {{ True }} sortAndCheck l {{ RET #true; True }} := ...

theorem sortAndCheckAdequate (l : List Int) (σ : State) :
    adequate .NotStuck (sortAndCheck l) σ
      (fun v _ => v = #true) := by
  apply heap_adequacy (GF := HeapLangS); intro _
  iapply sortAndCheck_spec <;> itrivial
\end{lstlisting}
\caption{Quicksort verification.}
\Description{Code formalizing a HeapLang quicksort implementation together with its Iris-Lean specification, proof script and adequacy theorem.}
\label{fig:quicksort}
\end{figure}

This section illustrates our port of HeapLang, the default language shipped with Iris, using the example in \Cref{fig:quicksort}.

\paragraph{Syntax}
Let us start by considering the implementation of the \leaninline|quicksort| and \leaninline|sortAndCheck| functions.
As in~\cref{fig:monotone-counter:lean}, we implement these functions with a dedicated ML-like syntax, invoked by the \leaninline|hl_val| and \leaninline|hl| macros.
The surface syntax reuses native Lean identifiers for its binders, although the macro emits a deeply embedded abstract syntax tree (AST), converting the identifiers to strings in the process.
To refer to Lean definitions inside a HeapLang program, one can use \leaninline|&|: \leaninline|quicksort| uses this to refer to the \leaninline|partition|, \leaninline|append|, and \leaninline|cons| functions defined earlier and omitted from the figure.\footnote{In Rocq, this corresponds to identifiers that are not in quotes.}
One can also use \leaninline|&| to embed more complex Lean expressions into a HeapLang program, such as in \leaninline|sortAndCheck|, which generates a HeapLang list from the given Lean list \leaninline|l| using the Lean \leaninline|makeList| function.
Afterwards, \leaninline|sortAndCheck| sorts the list using \leaninline|quicksort| and checks whether it is sorted.
\leaninline|sortAndCheck_spec| proves that this check always succeeds.

Iris-Lean also implements a linter to warn when a HeapLang variable is unbound.
This linter catches a class of typos, \eg writing \leaninline|cons| in place of \leaninline|&cons|.

\paragraph{HeapLang Proof Mode}
To reason about HeapLang programs, Iris-Lean provides tactics following Iris-Rocq. These are used in \leaninline|quicksort_spec| to prove that \leaninline|quicksort| returns a value \leaninline|l'| modeled by a sorted list \leaninline|ls'| (\leaninline|Pairwise LE.le ls'|) that is a permutation of the original list \leaninline|ls| (\leaninline|ls ~ ls'|).
The proof uses standard Iris tactics like \leaninline|iintro| to introduce Iris quantifiers and magic wands, and \leaninline|iloeb| to perform Löb induction.
In the same way as in Iris-Rocq, Iris-Lean also provides specialized tactics for symbolically executing HeapLang programs. For example, \leaninline|wp_pures| performs pure reduction steps, \leaninline|wp_load| executes a load instruction (\eg \leaninline|!x| in \leaninline|quicksort|), or \leaninline|wp_smart_apply| applies a lemma to a subexpression after potentially executing pure reductions.

\paragraph{Integration with Lean automation}
Iris-Lean can use the powerful automation provided by Lean in order to discharge pure side conditions, such as the final three lines of \leaninline|quicksort_spec|.
Here, one must prove that partitioning the list, sorting them individually, and then recombining them gives a sorted permutation of the input list.
Proving this requires combining different facts about filtering, sorting, and permutations.
We delegate this tedious reasoning to Lean's \leaninline|grind| tactic:
provided with the right facts from the standard library (\leaninline|pairwise_cons| and \leaninline|filter_append_perm|), \leaninline|grind| can discharge these proof obligations automatically.

\paragraph{Adequacy}
Just like its Rocq counterpart, Iris-Lean provides an adequacy result for HeapLang that turns an Iris-Lean verification of a HeapLang program into a closed statement about the operational semantics.
This adequacy theorem is exercised by \leaninline|sortAndCheckAdequate|:
this theorem proves that, for an arbitrary list \leaninline|l|, the \leaninline|sortAndCheck| function does not get stuck and, if it terminates, returns \leaninline|true|.
The theorem \leaninline|sortAndCheckAdequate| follows directly from \leaninline|sortAndCheck_spec| and the generic HeapLang adequacy theorem \leaninline|heap_adequacy|.

 \section{Differences between Iris-Lean and Iris-Rocq}
\label{sec:mechanization}

This section describes how Iris-Lean differs from Iris-Rocq, starting with the implementation of the Iris Proof Mode (\Cref{sec:metaprogramming}), then discussing how Iris-Lean replaces setoid equivalences with equality (\Cref{sec:setoid-to-eq}), and some engineering challenges (\Cref{sec:engineering}).

\subsection{The Iris Proof Mode}
\label{sec:metaprogramming}

The \emph{Iris Proof Mode} (IPM)~\citep{DBLP:conf/popl/KrebbersTB17} is what
makes interactive proofs within the Iris-Rocq framework powerful and practical, as it not
only abstracts away the basic properties of separation logics, such as the
associativity and commutativity of the separating conjunction, but also offers a
full set of tactics for proofs at the level of Iris entailments. Our
implementation of the IPM in Lean is based on the same theoretical foundations
as in Rocq, namely \emph{MoSeL}~\citep{mosel}.

\begin{figure}
\begin{lstlisting}[aboveskip=0pt,belowskip=0pt, basicstyle=\footnotesize\ttfamily]
    PROP : Type u
    inst : BI PROP
    φ : Prop
    h : φ
    P1 P2 Q : PROP
    ⊢
    ∗HP1 : P1
    □HP2 : P2
    ∗HPQ : ⌜φ⌝ -∗ P1 -∗ P2 -∗ Q
    ⊢ Q
\end{lstlisting}
\caption{An example of a proof state within the IPM.}
\Description{Iris Proof Mode display example showing the Lean context, followed by three named Iris hypotheses and then the goal.}
\label{fig:IPM-example}
\end{figure}

Within the IPM, the Iris context supplements the regular context in the proof state with \emph{intuitionistic} and \emph{spatial} hypotheses, whose names are prefixed with \leaninline|□| and \leaninline|∗|, respectively. IPM tactics are comparable to the built-in tactics in Lean. For example, \leaninline|iapply| and \leaninline|ispecialize| use magic wands similarly to how \leaninline|apply| and \leaninline|specialize| use implications in regular Lean proofs, albeit while potentially removing their hypotheses from the spatial context. In the proof state in \Cref{fig:IPM-example}, one can invoke \leaninline|ispecialize HPQ $$ ^h HP1 HP2| to discharge the premises in \leaninline|HPQ|; the spatial hypothesis \leaninline|HP1| is consumed while the pure hypothesis \leaninline|h| and intuitionistic hypothesis \leaninline|HP2| are not. The IPM also includes tactics for Iris-specific constructs, such as \leaninline|iloeb| for Löb induction and \leaninline|iinv| for opening invariants.

From the perspective of the user, the IPM tactics in Iris-Lean offer the same functionality as their counterparts in Iris-Rocq. However, the tactic implementations in Iris-Lean notably differ from Iris-Rocq in paradigm: whereas Iris-Rocq uses the Ltac tactic language to apply specialized lemmas to the proof goal, Iris-Lean tactics are implemented using Lean metaprogramming functions to manipulate the IPM hypotheses and goals as first-class objects.
The rest of this section describes this difference in more detail by highlighting (1) the representation of the proof state; (2) the internal tactic implementation paradigm; (3) Iris-Lean's custom typeclass synthesis algorithm; and (4) how these decisions surface in the user interface.

\paragraph{Proof mode representation}

\begin{figure}
\begin{lstlisting}[aboveskip=0pt,belowskip=0pt, basicstyle=\footnotesize\ttfamily]
  inductive Hyps {prop : Q(Type u)} (bi : Q(BI $prop)) :
      (e : Q($prop)) → Type where
    | emp (pf : $e =Q emp) : Hyps bi e
    | sep {elhs erhs}
      (lhs : Hyps bi elhs) (rhs : Hyps bi erhs)
      (pf : $e =Q iprop($elhs ∗ $erhs)) : Hyps bi e
    | hyp (name : Name) (p : Q(Bool)) (P : Q($prop))
      (pf : $e =Q iprop(□?$p $P)) : Hyps bi e
\end{lstlisting}
\caption{Definition of \leaninline|Hyps| (with minor simplifications).}
\Description{The definition of the inductive type Hyps, with constructors for the empty context, separating conjunctions and leaf nodes for hypotheses.}
\label{fig:Hyps}
\end{figure}

Iris-Lean uses a single inductive tree to represent the context, given by the type \leaninline|Hyps| shown in \Cref{fig:Hyps}. An value of \leaninline|Hyps| is either empty (\leaninline|emp|), a separating conjunction of two subtrees (\leaninline|sep|) or
a hypothesis (\leaninline|hyp|) consisting of its name (\leaninline|name|), a flag indicating whether it is intuitionistic or spatial (\leaninline|p|) and the proposition itself (\leaninline|P|).

A \leaninline|Hyps| value is dependently typed by the expression \leaninline|e| it represents. For example, the context in \Cref{fig:IPM-example} is represented by a value of type \leaninline|Hyps bi e|, where \leaninline|bi| is the BI instance for the logic and \leaninline|e| is the expression \leaninline|P1 ∗ □ P2 ∗ (⌜φ⌝ -∗ P1 -∗ P2 -∗ Q)|. Each node in \leaninline|Hyps| records in the field \leaninline|pf| that \leaninline|e| is equal to the represented proposition.

\begin{figure}
\begin{lstlisting}[aboveskip=0pt,belowskip=0pt, basicstyle=\footnotesize\ttfamily]
  def Hyps.add {prop : Q(Type u)}
    (bi : Q(BI $prop)) (name : Name) (p : Q(Bool))
    (P : Q($prop)) {e} (h : Hyps bi e) :
    (e' : Q($prop)) × Hyps bi e' ×
      Q(iprop($e ∗ □?$p $P ⊣⊢ $e'))
\end{lstlisting}
\caption{The type signature of \leaninline|Hyps.add|.}
\Description{Lean code giving the type signature of the function Hyps.add, which returns a new context with a new hypothesis added along with a proof relating the old and new contexts.}
\label{fig:Hyps.Add}
\end{figure}

To track these object-level propositions like \leaninline|e| in the metaprogram, Iris-Lean uses the Qq library~\citep{Qq}.
Qq provides \leaninline|Q(α)|: the type of Lean expressions (\leaninline|Lean.Expr|) of type \leaninline|α|.\footnote{Qq shares similarities with Mtac2~\citep{Mtac} in Rocq, but Qq is not tied to a specific monad, and its types are less strictly enforced.}
Typed expressions are useful when manipulating \leaninline|Hyps| using metaprogramming.
As an illustration, \leaninline|Hyps.add| (\Cref{fig:Hyps.Add}) is the function used for inserting a new hypothesis into the IPM context. In addition to the updated \leaninline|Hyps| value, it also returns a proof term asserting that the original context \leaninline|e| combined with the newly added hypothesis \leaninline|□?p P| corresponds to the new context \leaninline|e'|.
Thanks to Qq, the type of this proof term is part of the signature of \leaninline|Hyps.add|; the compiler checks that the tactic produces type-correct proof terms.
\leaninline|Hyps| provides a full interface of functions, meaning that the tactic implementations are agnostic about the low-level details of how hypotheses in the IPM context are managed.

\paragraph{Tactic implementations}
Context management is only one aspect of how the IPM in Iris-Lean differs from Iris-\rocq{}: rather than rewriting the proof goal in a sequence of steps, the tactic implementations in Iris-Lean are elaborators that run inside our domain-specific monad, \leaninline|ProofModeM|.
All Iris-Lean tactics essentially follow the same implementation paradigm:

\begin{enumerate}
  \item Use the function \leaninline|ProofModeM.runTactic| to parse the proof state into its components, such as the \leaninline|Hyps| value (\leaninline|hyps : Hyps bi e|), the conclusion of the entailment (\leaninline|goal|) and the metavariable for the goal (\leaninline|mvar|).
\item Introduce new subgoals for the user (\leaninline|addBIGoal|).
  \item Construct a proof of type \leaninline|Q($e ⊢ $goal)| for \leaninline|mvar| to close the original goal.
\end{enumerate}

As a concrete example, \cref{fig:isplit} shows the implementation of \leaninline|isplit|, which splits a conjunction (\leaninline|∧|) into two subgoals and keeps the entire IPM context for both. The function \leaninline|addBIGoal| performs two operations at once---it registers the new subgoal in the proof state while also immediately returning a term representing its proof. This is possible only because unsolved goals are ordinary values that can already be used within the metaprogram in Lean.
The \leaninline|ProofModeM| monad also allows us to provide various convenience functionality.
For example, the \leaninline|throwIPMError| macro allows us to raise consistent error messages across tactics by automatically including the current tactic name (which is tracked by the monad) in the error message.

\begin{figure}
\begin{lstlisting}[language=lean,aboveskip=0pt,belowskip=0pt, basicstyle=\footnotesize\ttfamily]
  theorem from_and_intro [BI PROP] {P Q A1 A2 : PROP}
      [FromAnd Q A1 A2] (h1 : P ⊢ A1) (h2 : P ⊢ A2) : P ⊢ Q

  elab "isplit " : tactic => do
    ProofModeM.runTactic `isplit λ mvar
        { prop, e, hyps, goal, .. } => do
      let A1 : Q($prop) ← mkFreshExprMVarQ prop
      let A2 : Q($prop) ← mkFreshExprMVarQ prop
      let some _ : Option Q(FromAnd $goal $A1 $A2) ←
          ProofModeM.trySynthInstanceQ
            q(FromAnd $goal $A1 $A2)
        | throwIPMError "{goal} is not a conjunction"
      let m1 : Q($e ⊢ $A1) ← addBIGoal hyps A1
      let m2 : Q($e ⊢ $A2) ← addBIGoal hyps A2
      mvar.assign q(from_and_intro (Q := $goal) $m1 $m2)
\end{lstlisting}
\caption{Implementation of \leaninline|isplit| as an elaborator using the theorem \leaninline|from_and_intro|.}
\Description{Lean code implementing the isplit tactic as an elaborator that synthesizes a FromAnd instance, registers two subgoals and assigns the resulting proof term.}
\label{fig:isplit}
\end{figure}

To inspect the shape of the goal, \leaninline|isplit|, following Iris-Rocq, uses the \leaninline|FromAnd goal A1 A2| typeclass, which searches for propositions \leaninline|A1| and \leaninline|A2| such that \leaninline|goal| corresponds to \leaninline|A1 ∧ A2|.
Typeclasses like \lstinline|FromAnd| let users extend \leaninline|isplit| to work on other connectives (\eg to split a pure goal $\ulcorner\varphi\land\psi\urcorner$ into $\ulcorner\varphi\urcorner$ and $\ulcorner\psi\urcorner$) just by adding new instances.
The \leaninline|isplit| tactic searches for a \leaninline|FromAnd| instance using the function \leaninline|ProofModeM.trySynthInstanceQ|, which we turn to next.

\paragraph{Custom typeclass synthesizer}

The extensibility of tactic functionality relies on the typeclass search mechanism.
It is critical that Iris-Lean includes a typeclass search mechanism consistent with Iris-Rocq, despite the differences in how Lean's built-in typeclass synthesizer operates.

In Iris-Rocq, \rocqinline|tc_solve| is used to invoke typeclass search, with \rocqinline|Hint Mode| declarations constraining the positions of the parameters that are allowed to be metavariables.\footnote{Unification variables or existential variables in \rocq{} parlance.}
Lean also provides a typeclass synthesis mechanism, but it does not allow instances to create or instantiate metavariables.
While this restriction enables Lean's synthesizer to cache the typeclass search results, the IPM relies on instances that create and instantiate metavariables (\eg for instantiating universal quantifiers with metavariables when applying a lemma), so the default synthesis procedure is not suitable.
We would also like a metavariable in an input position to fail one instance rather than aborting the entire search.
To address this gap, we implement a custom typeclass synthesizer specifically for IPM tactics.
This might sound like a daunting task, but since Lean's built-in typeclass synthesizer is implemented in Lean, we can reuse large parts of its functionality for the IPM synthesizer, including the complete infrastructure for declaring classes and instances.
Typeclasses intended for our IPM synthesizer just need to be labeled with the \leaninline|@[ipm_class]| annotation, and synthesis is invoked as \leaninline|ProofModeM.trySynthInstanceQ|.
The IPM synthesizer is implemented as a simple recursive search over the registered instances.
Backtracking is disabled by default and can be enabled per instance using the \leaninline|@[ipm_backtrack]| attribute. This keeps the proof search algorithm predictable and avoids accidental blow-ups from unintended backtracking.

Having full control of the synthesizer's implementation lets us smooth over some rough edges from Iris-\rocq{}.
In place of \rocqinline|Hint Mode| in Rocq, typeclasses in Lean use annotations in the signature to determine which parameters are inputs and which ones are outputs.
IPM classes offer fine-grained control over the synthesis direction.
As an example, \leaninline|IsOp| (\Cref{fig:IsOp}) is used by tactics such as \leaninline|icombine| and \leaninline|icases| to merge and split expressions.
Iris-Rocq needs \emph{three separate typeclasses} to deal with this, so that each can have its own \rocqinline|Hint Mode| configuration. Our synthesizer can express this using a single typeclass:
the direction \leaninline|d|---either \leaninline|.split| or \leaninline|.merge|---determines whether \leaninline|a| is split into \leaninline|b1| and \leaninline|b2| or whether \leaninline|b1| and \leaninline|b2| are merged, resulting in \leaninline|a|.
This is encoded using the \leaninline|semiOutParamIPM| annotation, whose first argument determines whether the parameter is an input or an output.
An instance can be applicable in both directions (\eg \leaninline|isOpFrac_half|) or only in a specific direction (\eg \leaninline|isOpFrac_merge| and \leaninline|isOpFrac_split|). The different priorities ensure that \leaninline|q.half + q.half| is split into \leaninline|q.half| and \leaninline|q.half|, but these are merged into \leaninline|q|.

\begin{figure}
\begin{lstlisting}[language=lean,aboveskip=0pt,belowskip=0pt, basicstyle=\footnotesize\ttfamily]
@[ipm_class]
class IsOp [CMRA α]
    (d : IsOp.Direction) (a : semiOutParamIPM d.toInOut α)
    (b1 : semiOutParamIPM d.toInOut.negate α)
    (b2 : semiOutParamIPM d.toInOut.negate α) where
  is_op : a = b1 • b2
instance (priority := low) isOpFrac_merge (q1 q2 : Qp) :
    IsOp .merge (q1 + q2) q1 q2
instance isOpFrac_half d (q : Qp) :
    IsOp d q q.half q.half
instance (priority := high) isOpFrac_split (q1 q2 : Qp) :
    IsOp .split (q1 + q2) q1 q2
\end{lstlisting}
\caption{The IPM typeclass \leaninline|IsOp| along with its instances for fractional permissions (\leaninline|Qp|).}
\Description{Lean code for the typeclass IsOp with a parameter indicating the direction, followed by three instances for fractional permissions at different priorities.}
\label{fig:IsOp}
\end{figure}

Finally, our IPM typeclass instances are not limited to the ordinary format of a declarative instance---it is also possible to define instances as Lean functions with the annotation \leaninline|@[ipm_tactic_instance]|, which corresponds to \rocqinline|Hint Extern| in Rocq. Such an instance is free to utilize arbitrary metaprogramming to build the synthesis result and can invoke synthesis recursively. The classes \leaninline|NatCancel| and \leaninline|TCSideCondition| in \Cref{fig:natCancel} are two prominent examples. The former is used for canceling common parts shared by the arithmetic expressions \leaninline|n| and \leaninline|m|, with the output parameter \leaninline|stuck| indicating whether it makes progress to cancel any expressions. What is implemented in Rocq using three typeclasses and over a dozen instances with delicate configuration of instance priorities can be implemented as a single function in Lean that programmatically performs the arithmetic manipulations. Meanwhile, \leaninline|TCSideCondition| uses tactics such as \leaninline|simp| for discharging proof obligations.

\begin{figure}
\begin{lstlisting}[language=lean,aboveskip=0pt,belowskip=0pt, basicstyle=\footnotesize\ttfamily]
@[ipm_class] class NatCancel (n m : Nat)
    (n' m' : outParam Nat) (stuck : outParam Bool) where
  nat_cancel : n' + m = n + m'
@[ipm_tactic_instance NatCancel _ _ _ _ _]
def instNatCancel : SynthTactic := λ e => do ...

@[ipm_class] class TCSideCondition (φ : Prop) : Prop where
  sidecondition : φ
@[ipm_tactic_instance TCSideCondition _]
def solveTCSideCondition : SynthTactic := fun e => do ...
\end{lstlisting}
\caption{IPM typeclasses with non-declarative instances.}
\Description{Lean code implementing the typeclasses NatCancel and TCSideCondition, each paired with a non-declarative instance defined as a metaprogram.}
\label{fig:natCancel}
\end{figure}

\paragraph{UI features}

The expressive metaprogramming infrastructure in Lean shows its value not only within the internal implementation details of the IPM but also in its user interface. The most visible advantage is that the IPM tactic syntax is handled directly by Lean's parser without the use of any string literals. Consequently, we can not only reuse syntax parsing machinery defined for \leaninline|rcases| and \leaninline|induction| for IPM tactics (\eg in \leaninline|icases| and \leaninline|iinduction|) but also to offer localized error messages (\eg for a specific part of a case destruction pattern).

\begin{figure}
\begin{lstlisting}[aboveskip=0pt,belowskip=0pt, basicstyle=\footnotesize\ttfamily,literate=
  {this}{this}4
  {apply}{{\color{Blue}apply}}5]
  Try this:
    [apply] irevert ^x ^h H
  Try this:
    [apply] irevert! ^x
\end{lstlisting}
\caption{Suggestions to correct the tactic usage.}
\Description{Rendering of two clickable ``Try this'' suggestions offered by the irevert tactic.}
\label{fig:irevert-example}
\end{figure}

We have also introduced features to specifically address differences between Lean and Rocq. For example, the tactic \leaninline|revert h| in Lean reverts not only the hypothesis \leaninline|h| but also other hypotheses that are forward-dependent on \leaninline|h|. Meanwhile, the tactic \rocqinline|iRevert| in Iris-Rocq fails when dependent hypotheses are not reverted in the correct order explicitly.
This is especially inconvenient in Lean, where hypotheses may be unnamed.
As a solution, we offer two variations of the tactic---\leaninline|irevert|, consistent with the Rocq tactic, and \leaninline|irevert!|, consistent with Lean's \leaninline|revert|, which automatically reverts dependent hypotheses. For example, suppose the Iris hypothesis \leaninline|H| references a pure Lean hypothesis \leaninline|h|, which in turn depends on the variable \leaninline|x|. Invoking \leaninline|irevert ^x|
raises an error since \leaninline|h| and \leaninline|H| depend on \leaninline|x|.
In this situation, the tactic interactively suggests corrected calls to \leaninline|irevert| or \leaninline|irevert!| as shown in \Cref{fig:irevert-example}, which the user can accept by clicking on \lstinline[language=lean,literate={apply}{{\color{Blue}apply}}5]|[apply]|. The same feature is available for other tactics that generalize hypotheses, such as \leaninline|iinduction|.

The IPM in Lean provides further UI features such as type information when hovering over terms and hypothesis names, as well as jumping to definitions, both in the tactic script and in the goal display.
Since the IPM is implemented in Lean like the standard Lean tactics, these features can be implemented by reusing standard Lean functionality also used by other tactics.
This is unlike Rocq, where standard tactics are usually implemented in OCaml and IPM tactics are implemented in Ltac.

\subsection{Replacing Setoid Equivalence by Equality}
\label{sec:setoid-to-eq}

\paragraph{Setoids}
\emph{Setoids}~\citep{irisjournal} are a key feature of the Iris-\rocq algebraic hierarchy: every OFE carrier comes with an equivalence relation (denoted $\equiv$) and step-indexed distances (denoted $\nequivB{n}$), which are related by the law \lstinline|equiv_dist|~: $x \equiv y \leftrightarrow \forall n,\, x \nequivB{n} y$.
The algebraic laws of Iris are only required to hold \emph{up to $\equiv$}, meaning users of Iris-\rocq are free to represent an OFE using any type whose propositional equality is \emph{finer} than their intended equivalence.
For example, an OFE of sets can be realized over the type \lstinline|list| by defining its setoid equivalence to be bi-inclusion; $\textrm{\lstinline|[a, a]|} \equiv \textrm{\lstinline|[a]|}$ even though $\textrm{\lstinline|[a, a]|} \ne \textrm{\lstinline|[a]|}$.

In \rocq, setoids are quite common:
\begin{itemize}
\item \rocq's type theory has no quotient types; setoids are the canonical workaround.
\item Some OFEs use setoids to avoid introducing axioms, \eg functional extensionality for function-space OFEs.
\end{itemize}
The complexity added by this extra congruence layer is handled in \rocq using \emph{generalized (setoid) rewriting}~\citep{DBLP:journals/jfrea/Sozeau09}.
By registering terms as being $\equiv$-preserving, the \rocqinline|rewrite| tactic is capable of manipulating terms up to equivalence during Iris proofs.
The price is an unavoidable body of congruence bookkeeping: \lstinline|Proper| instances for every operation (149 occurrences in {\lstinline|iris/algebra|} alone), \lstinline|_proper| lemmas, and the \lstinline|Leibniz|/\lstinline|leibniz_equiv| typeclass with a parallel family of \lstinline|_L| lemmas (46 occurrences under {\lstinline|iris/algebra|}) for the special case where $\equiv$ does coincide with $=$.

Lean 4 has no built-in counterpart of generalized rewriting: \lstinline|rw|/\lstinline|simp| rewrite with $=$ (and $\leftrightarrow$) only.
A faithful port of the setoid design leaves the user without its main rewriting tactic, turning every rewrite up to $\equiv$ into a chain of transitivity steps and manually applied congruence lemmas.
On the other hand, Lean comes with different technical and social norms.
Quotient types (\lstinline|Quot|) are a kernel primitive with a definitional computation rule, and the principles avoided by Iris-\rocq (functional extensionality, propositional extensionality, and choice) are freely assumed across Lean's standard library.\footnote{Iris-\rocq{} may be capable of abandoning setoids by adopting similar axioms, but \rocq{} libraries are generally expected to avoid doing so. }
This suggests a different design: require $\equiv$ to coincide with $=$ (in Iris-\rocq nomenclature this amounts to all OFEs being \emph{Leibniz}) and use quotient types when this property does not hold on the nose.

\paragraph{Leibniz OFEs}

\begin{figure}
\begin{lstlisting}[basicstyle=\footnotesize\ttfamily]
class OFE (α : Type _) where
  Dist : Nat → α → α → Prop
  dist_eqv : Equivalence (Dist n)
  eq_dist : x = y ↔ ∀ n, Dist n x y
  dist_lt : Dist n x y → m < n → Dist m x y
\end{lstlisting}
\caption{OFE definition in Iris-Lean.}
\Description{Lean code implementing the OFE typeclass, with fields for the distance family, its equivalence property, its relation to equality and downward closure.}
\label{fig:ofe-def}
\end{figure}

The Lean \lstinline|OFE| class (\cref{fig:ofe-def}) keeps only the distances and states the former \lstinline|equiv_dist| law directly against equality.
The field \lstinline|eq_dist| plays two roles: the right-to-left implication is the \emph{Leibniz} requirement (elements at distance $n$ for every $n$ are equal); the left-to-right direction is the usual compatibility of equality with distance.
We still have $\nequivB{n}$ as a family of relations. Non-expansiveness, contractivity, the later modality ($\later$), COFE limits, and the America--Rutten domain equation solver are the same.

The construction where \rocq's equivalence is most visibly \emph{not} equality is the agreement OFE/CMRA: a record of a non-empty list \lstinline|agree_car : list A| with distance
\begin{equation*}
    x \nequivB{n} y := (\forall a \in x,\, \exists b \in y,\, a \nequivB{n} b) \wedge
    (\forall b \in y,\, \exists a \in x,\, a \nequivB{n} b).
\end{equation*}
The corresponding equivalence $x \equiv y := \forall n,\, x \nequivB{n} y$ identifies duplicated and reordered lists, and must admit rules such as \lstinline|agree_idemp|~: $a \cdot a \equiv a$.
\Cref{fig:agree-def} shows the two-step construction of \lstinline|Agree| in Iris-Lean.
We first replicate the definition in Iris-\rocq as \lstinline|Agree.Raw|, and then quotient it by the required setoid equivalence.

\begin{figure}
\begin{lstlisting}[basicstyle=\footnotesize\ttfamily]
structure Agree.Raw α where
  car : List α
  not_nil : car ≠ []
def SameElems (x y : Agree.Raw α) : Prop :=
  (∀ a ∈ x.car, a ∈ y.car) ∧ (∀ b ∈ y.car, b ∈ x.car)
def Agree (α : Type u) : Type u :=
  Quotient ⟨SameElems, ...⟩
\end{lstlisting}
\caption{Constructing \lstinline|Agree| as a quotient.}
\Description{Lean code implementing the type Agree as a quotient of a raw non-empty-list structure by a same-elements relation.}
\label{fig:agree-def}
\end{figure}

\begin{figure}
\begin{lstlisting}[basicstyle=\footnotesize\ttfamily]
def OFE.ofQuotient {s : Setoid X}
    (dist : Nat → X → X → Prop)
    (dist_eqv : Equivalence (dist n))
    (dist_lt : dist n x y → m < n → dist m x y)
    (coincide : ∀ x y, x ≈ y ↔ ∀ n, dist n x y) :
    OFE (Quotient s)
\end{lstlisting}
\caption{The OFE structure on quotients.}
\Description{Lean code showing the type signature of OFE.ofQuotient, which builds an OFE on a quotient from a distance family and a proof that it agrees with the setoid relation.}
\label{fig:quot-ofe}
\end{figure}

We recover the OFE structure on \lstinline|Agree| as an instance of a generic quotient OFE presented in~\cref{fig:quot-ofe}, setting the setoid \lstinline|s| to be \lstinline|SameElems|, and defining \lstinline|Raw.dist| as above:

\begin{minipage}{\linewidth}
\begin{lstlisting}[basicstyle=\footnotesize\ttfamily]
def Agree.Raw.dist (n : Nat) (x y : Raw α) : Prop :=
  (∀ a ∈ x.car, ∃ b ∈ y.car, a ≡{n}≡ b) ∧
  (∀ b ∈ y.car, ∃ a ∈ x.car, a ≡{n}≡ b)
instance [OFE α] : OFE (Agree α) :=
  OFE.ofQuotient Agree.Raw.dist ... sameElems_iff_dist
\end{lstlisting}
\end{minipage}

To construct the quotient OFE, we must supply a proof that the setoid relation \lstinline|SameElems| coincides with the intersection of distances (\lstinline|sameElems_iff_dist|).
The forward direction \lstinline|SameElems x y → ∀ n, Raw.dist n x y| is trivial.
For the reverse direction, we must show that agreement at every distance implies \lstinline|SameElems|.
Let $a \in x$. At every index $n$ there exists a witness $b \in y$ such that $a \nequivB{n} b$.
Since $y$ is finite and there are infinitely many indices, a classical pigeonhole argument returns a single $b \in y$ with $a \nequivB{n} b$ for arbitrarily large $n$.
Then, downwards closure (\lstinline|dist_lt|) implies that $a \nequivB{n} b$ at every index, and by the Leibniz property of $\alpha$ we get $a = b$.
Note that this argument relies on $\alpha$ itself being Leibniz, though with our change to OFE, this is always true. 

In the quotient, algebra laws are proved on \lstinline|Raw| since distance lemmas can be converted into equalities on the quotient.
For instance, \leaninline|Raw.idemp : ∀ n, x • x ≡{n}≡ x| can be promoted to \leaninline|op_idemp : x • x = x| by \leaninline|eq_dist|.

We close this section by discussing another important OFE which fails to be Leibniz for a different reason.
Monotone predicates over step-indices and resources (\lstinline|uPred|) underlies the model of the Iris base logic.
The setoid equivalence on \lstinline|uPred| identifies predicates that are logically equivalent over the set of valid index-resource pairs, \ie \lstinline|P ≡ Q| when \lstinline|∀ n x, ✓{n} x → (P n x ↔ Q n x)|.
This construction is not Leibniz in \rocq for several reasons (\eg all invalid elements are identified with each other).

In Iris-Lean, we modify the definition of \lstinline|uPred M| to quantify over elements that are \emph{valid by construction}:

\begin{minipage}{\linewidth}
\begin{lstlisting}[basicstyle=\footnotesize\ttfamily]
structure UPred (M : Type _) [UCMRA M] where
  holds : (n : Nat) → {x : M // ✓{n} x} → Prop
  mono : ...
\end{lstlisting}
\end{minipage}
Now \lstinline|funext|, \lstinline|propext| and proof irrelevance imply that \lstinline|UPred| is Leibniz.
Since \lstinline|uPred M| is the model of separation logic resources, this change percolates upwards.
Namely, in Iris-Lean, the equality of BI propositions also coincides with bi-entailment.
As such, logical rules like $P \sep Q \provesIff Q \sep P$ turn into $P \sep Q = Q \sep P$ via \leaninline|BiEntails.to_eq|, so \lstinline|rw| and \lstinline|simp| can rewrite by them anywhere, including under binders and in dependent positions, where \rocq's setoid rewriting fails.

\paragraph{Benefits and costs}
With equality in place of $\equiv$, Lean's rewriting tools apply directly to the algebra laws, and equational laws can be added to \lstinline|simp| sets (unit laws, idempotence, \etc{}), so \lstinline|simp| can normalize terms.
\rocq's \rocqinline|autorewrite| offers an analogue for setoid equivalence, but every step is still a generalized rewrite, so it inherits the bookkeeping costs that the setoid-free design removes.
These conveniences are not free: each non-Leibniz construction needs lifting boilerplate, around 90 lines for \lstinline|Agree|.
We also note that Iris-Lean does not yet support rewriting by single entailments $\vdash$ or the step-indexed distances $\nequivB{n}$.
Mathlib's \lstinline|grw| and K\"{o}nig's original \lstinline|rw'| tactic \citep{koenig22} show promise, but we have yet to identify a technique that is on par with Iris-\rocq{}.

\subsection{Engineering Challenges}
\label{sec:engineering}
Some features of Iris-Rocq required us to take alternate routes and come up with new ways to express the constructions in Iris.
The two differences we cover in this section are \rocq's cumulative universes and canonical structures.
In the following we describe how Iris-\rocq{} uses each, as well as their replacements in Iris-Lean.

\paragraph{Universe cumulativity and abstract quantifiers}
In Iris-\rocq, quantifiers are stated using a fixed universe \lstinline|Quant|:

\begin{minipage}{\linewidth}
\begin{lstlisting}[language=coq,frame=none,basicstyle=\footnotesize\ttfamily]
bi_forall : ∀ A : Type@{Quant}, (A → bi_car) → bi_car
\end{lstlisting}
\end{minipage}
This single universe is enough since \rocq's universes are cumulative~\citep{DBLP:conf/itp/SozeauT14}: every type at a level below \lstinline|Quant| also lives in \lstinline|Type@{Quant}|.
Lean's universes are not cumulative, so an identical design would only allow quantification over types at exactly one level.
Parameterizing by the level is not enough either: structure fields cannot bind universe variables, so \lstinline|BIBase| instances at different levels would be unrelated.

Instead, Iris-Lean states quantification over \emph{sets} of propositions as primitive:
\leaninline|sForall : (PROP → Prop) → PROP|.
No universe appears in its type, by impredicativity of \lstinline|Prop|, and the usual indexed quantifiers are defined as the infimum and supremum of the image of the body:

\begin{minipage}{\linewidth}
\begin{lstlisting}[language=lean,basicstyle=\footnotesize\ttfamily]
def «forall» {α : Sort _} (P : α → PROP) : PROP :=
  sForall fun p => ∃ a, P a = p
\end{lstlisting}
\end{minipage}

The predicate \leaninline|fun p => ∃ a, P a = p| lands in \lstinline|Prop| no matter which universe $\alpha$ lives in, so \lstinline|«forall»| ranges over all of \lstinline|Sort _|, including \lstinline|PROP| itself.
The \rocq interface offers less: each use of a quantifier adds a global constraint on \lstinline|Quant|, and conflicting constraints surface as universe inconsistencies.
The entailment rules are stated once for \lstinline|sForall|/\lstinline|sExists| and derived for the quantifiers.

It is notable that, following on from this work, a merge request applying the same approach to Iris-Rocq was created by one of its maintainers~\citep{iris-merge-request-quant-universe}.
If there comes a situation where the Iris-Rocq developers seek to remove that additional universe constraint, our work in Iris-Lean demonstrates that the core constructions of Iris generalize to the set-based construction.

\paragraph{Bundled Hierarchies and Canonical Structures}
The core theory of Iris is built atop a hierarchy of algebraic structures, and inheritance is prevalent throughout.
One important decision when implementing an algebraic hierarchy in a dependently typed proof assistant is the degree to which structures are bundled, \ie which types, terms, and properties belong as fields of data structures versus parameters of them.
The bundling discipline of an algebraic hierarchy has a significant impact on which kinds of structure inference are possible and efficient.

Iris-Rocq implements this using a sophisticated combination of semi-bundled typeclasses and bundled canonical structures, arising in part from performance issues with Rocq's typeclass synthesis algorithm early on in the development of Iris~\citep{jung19typeclasses}.
While this mixed approach has been successful in Rocq, it is not without limitations.
A systematic approach for building bundled hierarchies in Rocq has been developed as part of MathComp and condensed into a general tool called Hierarchy Builder \citep{cohen_et_al:LIPIcs.FSCD.2020.34}.
However, Hierarchy Builder does not support embedding typeclass fields in the algebra, a feature that Iris requires to support generalized rewriting.
This makes Hierarchy Builder not suitable for Iris.
To this day, there are certain extensions to the Iris hierarchy which cannot be added in a modular way: Iris-\rocq{} reverted the \rocqinline|Duplicable| typeclass shortly after it had been merged for this reason~\citep{iris-issue-duplicable}.

Lean has excellent built-in support for hierarchies based on semi-bundled typeclasses, and Mathlib demonstrates that this approach scales well beyond the needs of Iris, so Iris-Lean fully embraces a typeclass-based hierarchy.
This makes the hierarchy easy to extend while also integrating well with Lean's existing libraries.
For instance, in \cref{subsec:prob} we will encounter a type that carries both an Iris language and a Mathlib \lstinline|MeasurableSpace| instance.
Here, adding a typeclass instance for each structure is vastly easier than in the bundled approach, wherein we would need a new layer in the hierarchy combining the properties of both structures.

We found that fully bundled structures were only really necessary in the definition of the \lstinline|IProp| model.
There, the construction requires one to prove that a fixed point over a list of resource functors induces a function from indices to resources.
The well-formedness of this function requires the resource structure associated to each functor to be equal: this is not provable using unbundled typeclasses, so we locally bundle a single canonical choice for the resource structure before taking the fixed point.
Nevertheless, we were able to abstract over this bundling with a typeclass-based interface, so clients of \lstinline|IProp| can avoid the bundled terms entirely.

 \section{Applications}
\label{sec:evaluation}

In this section, we present some new developments using Iris-Lean
related to motivations (1), (2) and (3) from the introduction: metaprogramming, automation and Mathlib.

\subsection{Generalizing Probabilistic Program Logics}
\label{subsec:prob}

A benefit of developing an Iris project in Lean is access to its library of mechanized mathematics.
As a case study, we will show how the combination of Iris-Lean and Mathlib can already address a gap in the Iris literature.
In particular, we develop a \emph{Continuous Probabilistic Program Logic}: a program logic for
languages which generate and manipulate random real numbers.
Languages of this sort are attractive targets for verification as they serve as idealized models for processes in cryptography and differential privacy.

Unfortunately, they have proven to be challenging to mechanize in Iris.
At the time of writing, none of the probabilistic Iris-based logics~\citep{DBLP:journals/pacmpl/GregersenAHTB24, gregersen2024refinement, DBLP:journals/pacmpl/0001HMLGTB24, haselwarter2024tachis, haselwarter2025approxis, li2025coneris, marionneau2026cpp, haselwarter2026dp, lohse2024irisexpectedcostanalysis, lohse2026stepsprobabilisticirisharmonizing} are capable of handling languages with a continuous semantics.
The closest is \emph{Continuous Eris} (CE)~\citep{demedeiros2026ce}; however, the CE logic has a number of limitations.
CE can only verify samplers from particularly well-behaved distributions,\footnote{Distributions with countably piecewise-continuous probability mass functions.} only against one implementation of real sampling, and its technique based on time receipts~\citep{mevel2019timereceipts} prevents it from establishing the probabilistic termination of its samplers as in Total Eris (TE).
Each of these issues stems from the same place: CE, like all other Iris-based probabilistic logics to date, avoids measure-theoretic probability in \rocq{}.

We present \emph{Continuous Total Eris} (CTE), a continuous generalization of Total Eris which addresses the aforementioned limitations of \emph{Continuous Eris}.
To our knowledge, this is
the first program logic capable of formally verifying properties of higher-order, stateful, continuous probabilistic programs using conventional measure theory.

\paragraph{Generalizing Total Eris}

Generalizing TE to CTE changes little about the program logic---the difficulty lies in altering the underlying operational semantics to use more general definitions from measure theory.
Two of the main challenges are (1) equipping the type of program expressions with a non-trivial \lstinline|MeasurableSpace|, and (2) generalizing countable sums in the program semantics to \emph{Lebesgue integrals}.
Both pose issues for the Rocq implementation of TE.

Our development of CTE began with a one-to-one port of TE to Iris-Lean.
This port was supported by LLMs: a multi-agent system completed an initial version of this port within a few days.
We then addressed challenge 1: equipping the type of program expressions with the structure of a
\lstinline|MeasurableSpace|.
The space itself is induced by the projection maps from expressions into their sub-expressions; this is a standard construction in measure theory and is implemented using Mathlib's \lstinline|MeasurableSpace.generateFrom|.
Changing the underlying measure space means that nearly every theorem in TE is broken. However, here the unbundled typeclass hierarchy of Mathlib comes to the rescue.
The \lstinline|MeasurableSpace| on program expressions in CTE is parameterized by the \lstinline|MeasurableSpace| on the type of real numbers \lstinline|rT|, and we proved that if the real numbers are equipped with a discrete \lstinline|MeasurableSpace| then the remainder of the development collapses to the discrete case as well.
Therefore, by postulating the discreteness of \lstinline{rT} on a per-declaration level, we obtained an intermediate version of TE that has measure-theoretic foundations but a discrete program logic.

Addressing challenge 2 involved generalizing the proofs and definitions of TE in order to remove these discreteness hypotheses.
Lean assisted us here as well: we wrote temporary linters and attributes to track the dependencies between definitions and lemmas as they were gradually generalized, preserving the build at all stages, and letting us explore alternate generalizations by tweaking high-level statements in the program logic.
Through personal correspondence with the authors of CE, we learned that the front-loading of challenges 1 and 2 stalled their initial attempts at making the same generalization: CTE did not suffer from these issues.

\paragraph{Corollaries of CTE}

We have explained how Lean was instrumental in generalizing TE to CTE.
Now we will outline three ways that access to Mathlib makes our logic stronger.

First, we resolve the open question in Section 7 of CE, where the authors propose a proof rule for sampling uniformly over $[0, 1]$ while avoiding all rational numbers:
\begin{equation*}
\thoare{\,}{\langkw{urand}}{r : \mathbb{R}.\, r \in [0,1] \wedge r \not\in \mathbb{Q}}
\end{equation*}
The suggested proof using CE's \emph{error credit technique} amounts to showing that the integral of the indicator function of $\mathbb{Q} \cap [0,1]$ is arbitrarily small.
The authors note that CE cannot mechanize this proof as this indicator function is not Riemann integrable.
This is not an issue in CTE: the Mathlib theorem establishing that the Lebesgue measure of any countable subset of $\mathbb{R}$ is zero allows us to complete the proof.

For our second example, we verify an algorithm by \citet{karney16normal} for sampling exactly from the unit normal distribution.
This is the main case study of CE, and we largely adapt their proofs. However, as our logic is total, we replace L\"{o}b induction with the \emph{credit induction} technique from TE.
As a result, our theorems are stronger: we verify both that the algorithm is correct \emph{and} that it terminates with probability one.
This proof is impossible in CE, as the \emph{time receipts} technique they use to avoid measure theory is unsound in a total logic.

Finally, we prove a suite of concentration bounds for our implementation of the unit normal sampler.
A concentration bound establishes the probability that a random sample falls outside a certain range, and can be used to analyze the accuracy of differentially private programs.
\Cref{eqn:concentration} is the general formulation of a concentration bound in CTE; it is read as \emph{``For any $t > 0$, the probability that the unit normal sampler returns a value outside $[-t, t]$ is at most $B(t)$''.}\footnote{The \emph{error credit} resource $\upto{B(t)}$ in the precondition is the Eris technique for specifying the probability that the postcondition does not hold.}
\begin{equation}
\label{eqn:concentration}
\forall t > 0,\, \thoare{\upto{B(t)}}{\langkw{unitNormal}}{r : \mathbb{R}.\, \vert r \vert < t}
\end{equation}

Section 5.2 of CE proves a simple concentration bound for their Laplace sampler, but their method of direct calculation is a special case that does not work for the normal distribution.
Instead, concentration bounds on the normal distribution typically follow from Markov's inequality ($B(t) = \sqrt{2/\pi}\,t^{-1}$), Chebyshev's inequality ($B(t) = t^{-2}$), Chernoff's inequality ($B(t) = e^{-t^2/2}$), or Mills' inequality ($B(t) = \sqrt{2/\pi} e^{-t^2/2} t^{-1}$), to name a few.
In CTE, we establish these four concentration bounds as instances of~\cref{eqn:concentration}.
The bulk of the work is done by Mathlib, which includes a substantial development of classical probability theory and allows us to establish all four rules in under 300 lines.

\subsection{Wander: Proof Mode Automation}

In \cref{sec:heaplang}, we saw how \leaninline|grind| can automate \emph{pure} parts of an Iris proof.
This raises the question of whether it is also possible to automate the parts that pertain to separation logic.
As it turns out, Iris proofs very often follow familiar patterns
    and are therefore amenable to \emph{goal-directed search},
    with each proof step chosen without backtracking according to the shape of the goal.
In Iris-\rocq, these regularities are exploited by Diaframe~\citep{mulder2022diaframe} and Lithium~\citep{RefinedC} to implement
    a set of tactics which enable the largely automatic verification of many programs and data structures in Iris.

\emph{Wander} is an Iris-Lean tactic designed to provide the same automation as Diaframe, but implemented
    using Lean's metaprogramming APIs. Doing so presents us with some benefits.
One such benefit is the leverage of other tactics' implementations in our development, such as \leaninline|simp|.
Not only are \leaninline|simp|'s internals used to replace Diaframe's custom-built simplifier, but its more
    mature implementation can be used to subsume other Diaframe components.
Another benefit is access to Lean's profiling and tracing infrastructure, which proves to be a great aid in the continued
    development of Wander.

The Wander tactic gives access to an interactive environment where users are able to direct Wander's behaviour
through a DSL of \emph{actions}:
    \texttt{step} takes a single (goal-directed) proof step,
    \texttt{continue} keeps taking such steps until stuck, and
    \texttt{finish} ensures these steps close the Iris goal, reporting an error otherwise.
Since Wander is only able to handle separation logic goals, pure goals encountered are stored internally as side conditions.
Side conditions can be dispatched by external tactics using the \texttt{sideconditions with} action.
In \Cref{fig:wander_quicksort}, we show how the quicksort example from \cref{sec:heaplang} can be solved using Wander.
With the right configuration, Wander is able to automate all of the Iris reasoning, leaving the pure goals to be handled by \leaninline|grind|.

\begin{figure}
\begin{lstlisting}[language=lean,aboveskip=0pt,belowskip=0pt, basicstyle=\footnotesize\ttfamily]
theorem quicksort_spec l ls :
    {{ isList (GF := GF) l ls }}
      hl(&quicksort &l)
    {{ l' ls', RET l'; isList l' ls' ∗
      ⌜Pairwise LE.le ls'⌝ ∗ ⌜ls ~ ls'⌝ }} := by
  iloeb as IH generalizing ^l ^ls
  unfold quicksort
  wander =>
    finish
    sideconditions with
    · grind [filter_append_perm]
    · rename List Int => l2; rename Int => h
      have : l2.all (h < ·) := by grind
      grind [pairwise_cons]
\end{lstlisting}
\caption{Quicksort verification with Wander.}
\Description{Lean code for the quicksort proof rewritten with the Wander tactic, where reasoning is automated, and only side conditions are manually discharged.}
\label{fig:wander_quicksort}
\end{figure}

\subsection{Automating Fixed Points}

The logic of Iris offers a rich supply of fixed point constructors: there are combinators for least and greatest fixed points of monotone functions as well as fixed points of contractive functions based on the ``later'' modality.
These constructions are used in many definitions, including the weakest precondition and total weakest precondition connectives.
Definitions using fixed points are typically written in three stages:
(1) first the user specifies the recursion template,
(2) then the user proves monotonicity or contractivity, and
(3) finally the fixed point combinator is used to complete the desired definition.
Going through these three stages makes for an awkward user experience.
For example, when defining monotone fixed points with multiple arguments, the recursion template \emph{must} be uncurried before the fixed point combinator is applied, because the monotonicity typeclass is limited to unary predicates, an implementation detail which must be known a priori.

In Iris-Lean we provide a custom suite of elaborators for defining Iris fixed points in a style similar to how recursive functions and inductive relations are defined in Lean, circumventing the awkward boilerplate.
Consider this example adapted from \citet{krebbers2025inductive}:

\begin{minipage}{\linewidth}
\begin{lstlisting}[basicstyle=\footnotesize\ttfamily]
ifix isDelList (l : Loc) (vs : List Val) :
    IProp GF := iprop^
  (l ↦ NIL ∗ ⌜vs = []⌝) ∨
  (∃ tl v vs', l ↦ CONS v tl ∗
    isDelList tl vs' ∗ ⌜vs = v :: vs'⌝) ∨
  (∃ tl, l ↦ DEL tl ∗ isDelList tl vs)
\end{lstlisting}
\end{minipage}

The predicate \leaninline|isDelList l vs| asserts that the location \leaninline|l| is a list, possibly including \leaninline|DEL| nodes representing deleted values, and containing the values \leaninline|vs|.
Note that the recursion is non-structural and the fixed point combinator is implicit.
Here, our \lstinline|ifix| elaborator takes the syntax of the recursive definition and handles stages (1), (2), and (3) of the boilerplate described above automatically.
In this example, the keyword \leaninline|ifix| tells the elaborator to take the least fixed point.
Likewise, one can write \leaninline|icofix| for the greatest fixed point, or \leaninline|iguarded| for the contractive fixed point.

For stage (2), it is necessary to ensure monotonicity or contractivity of the recursion template.
Using typeclass search, together with additional automation for purposes including uncurrying and case splitting on matches, our approach is able to automatically derive these conditions across most typical use cases. Should the automation fail, the user may also manually provide the required proof.

We also define an elaborator for specifying inductive relations within Iris, mirroring how one defines inductive relations in Lean.
With this notation, the above example can be written as:

\begin{minipage}{\linewidth}
\begin{lstlisting}[basicstyle=\footnotesize\ttfamily]
iinductive isDelList : Loc → List Val → IProp GF where
| nil (l : Loc) : l ↦ NIL -∗ isDelList l []
| cons (l tl : Loc) (v : Val) (vs : List Val) :
    l ↦ CONS v tl -∗
    isDelList tl vs -∗ isDelList l (v :: vs)
| del (l tl : Loc) (vs : List Val) :
    l ↦ DEL tl -∗ isDelList tl vs -∗ isDelList l vs
\end{lstlisting}
\end{minipage}

This command is similar to the \rocqinline|Iris Inductive| command of \citet{krebbers2025inductive} in Rocq, implemented using Elpi~\citep{Elpi}.
While Elpi provides convenient facilities to construct definitions, it can be difficult to integrate with the proof mode tactics written in Ltac.
For this reason, \citeauthor{krebbers2025inductive} reimplemented parts of the Iris Proof Mode in Elpi.
This is unnecessary in our case as IPM tactics are implemented within the same metaprogramming framework.

\subsection{Performance}

\begin{figure*}[t]
    \centering
    \begin{subfigure}[t]{0.48\textwidth}
        \begin{tikzpicture}
\begin{axis}[
    xlabel={Problem size (number of hypotheses)},
    ylabel={Time (s)},
    xmin=10,
    xmax=120,
    xtick distance=10,
    ymin=0,
    ymax=2.5,
    ytick distance=0.5,
    legend style={
        at={(0.5,-0.2)},
        anchor=north,
        legend columns=2,
    },
    grid=major,
    width=0.96\textwidth,
    height=6.6cm,
]

\addplot[
    mark=square*,
    mark size=1pt,
    dashed,
    color=Blue
] coordinates {
    (10,0.0153503)
    (15,0.0311489)
    (20,0.0537750)
    (25,0.0769551)
    (30,0.0871680)
    (35,0.1222874)
    (40,0.1628280)
    (45,0.2084706)
    (50,0.2670712)
    (55,0.3263681)
    (60,0.3972512)
    (65,0.4847846)
    (70,0.5734070)
    (75,0.6723260)
    (80,0.7931205)
    (85,0.9078589)
    (90,1.0435765)
    (95,1.2006087)
    (100,1.3621990)
    (105,1.5642772)
    (110,1.7381571)
    (115,1.9800657)
    (120,2.2345816)
};
\addlegendentry{Lean (forward order)}

\addplot[
    mark=*,
    mark size=1pt,
    dashed,
    color=Orange
] coordinates {
    (10,0.041)
    (15,0.088)
    (20,0.170)
    (25,0.300)
    (30,0.496)
    (35,0.798)
    (40,1.161)
    (45,1.663)
    (50,2.394)
};
\addlegendentry{Rocq (forward order)}

\addplot[
    mark=square*,
    mark size=1pt,
    color=Blue
] coordinates {
    (10,0.0038746)
    (15,0.0056357)
    (20,0.0073170)
    (25,0.0093297)
    (30,0.0115553)
    (35,0.0135837)
    (40,0.0154734)
    (45,0.0174922)
    (50,0.0198342)
    (55,0.0202983)
    (60,0.0216080)
    (65,0.0229784)
    (70,0.0252568)
    (75,0.0269999)
    (80,0.0294588)
    (85,0.0311978)
    (90,0.0327233)
    (95,0.0350651)
    (100,0.0368105)
    (105,0.0392717)
    (110,0.0413998)
    (115,0.0433095)
    (120,0.0463230)
};
\addlegendentry{Lean (reverse order)}

\addplot[
    mark=*,
    mark size=1pt,
    color=Orange
] coordinates {
    (10,0.007)
    (15,0.013)
    (20,0.022)
    (25,0.030)
    (30,0.043)
    (35,0.057)
    (40,0.076)
    (45,0.101)
    (50,0.146)
    (55,0.169)
    (60,0.201)
    (65,0.245)
    (70,0.295)
    (75,0.360)
    (80,0.420)
    (85,0.539)
    (90,0.566)
    (95,0.700)
    (100,0.760)
    (105,0.873)
    (110,1.021)
    (115,1.167)
    (120,1.256)
};
\addlegendentry{Rocq (reverse order)}

\end{axis}
\end{tikzpicture}         \caption{Time comparison of framing.}
        \Description{Line chart showing the time for framing in seconds, with the number of hypotheses from 10 to 120. For each of Iris-Lean and Iris-Rocq, there are two lines, one for the forward order and another for the reverse order. The lines for Iris-Rocq grow faster in both cases.}
        \label{fig:FrameBenchmark}
    \end{subfigure}
    \hfill
    \begin{subfigure}[t]{0.48\textwidth}
        \begin{tikzpicture}
\begin{axis}[
    xlabel={Problem size (list length)},
    ylabel={Time (s)},
    legend style={
        at={(0.5,-0.2)},
        anchor=north,
        legend columns=1,
    },
    grid=major,
    xmin=10,
    xtick distance=10,
    ytick distance=2,
    xmax=100,
    ymin=0,
    ymax=12,
    font=\small,
    width=0.96\textwidth,
    height=6.6cm,
]
\addplot[
    mark=square*,
    mark size=1pt,
    color=Blue
] coordinates {
    (10,0.085717)
    (15,0.122326)
    (20,0.168491)
    (25,0.215098)
    (30,0.266678)
    (35,0.316992)
    (40,0.367244)
    (45,0.415737)
    (50,0.469983)
    (55,0.523937)
    (60,0.579928)
    (65,0.636551)
    (70,0.697336)
    (75,0.749254)
    (80,0.812241)
    (85,0.883284)
    (90,0.950262)
    (95,0.998943)
    (100,1.059323)
};
\addlegendentry{Lean}

\addplot[
    mark=*,
    mark size=1pt,
    color=Orange
] coordinates {
    (10,0.279)
    (15,0.471)
    (20,0.736)
    (25,1.076)
    (30,1.377)
    (35,1.784)
    (40,2.246)
    (45,2.689)
    (50,3.277)
    (55,3.845)
    (60,4.429)
    (65,5.179)
    (70,5.866)
    (75,6.604)
    (80,7.473)
    (85,8.343)
    (90,9.249)
    (95,10.099)
    (100,11.235)
};
\addlegendentry{Rocq}

\end{axis}
\end{tikzpicture}
         \caption{Time comparison of HeapLang verification.}
        \Description{Line chart of HeapLang verification time in seconds against list length from 10 to 100, showing roughly linear growth for Iris-Lean and superlinear growth for Iris-Rocq.}
        \label{fig:HeapLangBenchmark}
    \end{subfigure}
    \vspace{1em}
    \caption{Benchmark results.}
    \Description{Two line charts comparing the runtime using Iris-Lean and Iris-Rocq as the problem size grows.}
\end{figure*}
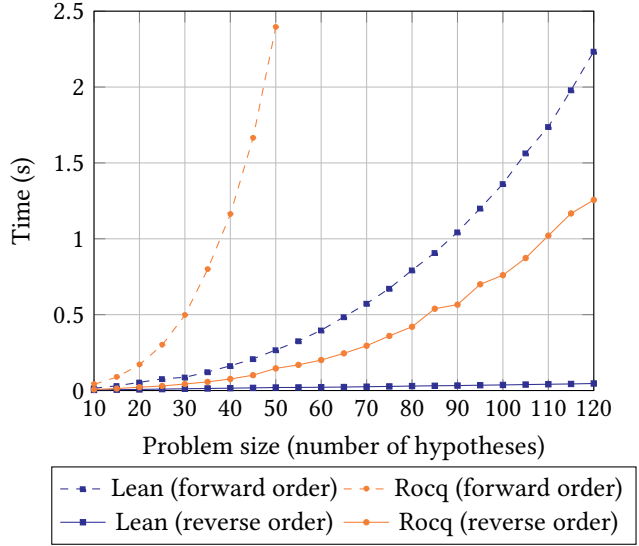
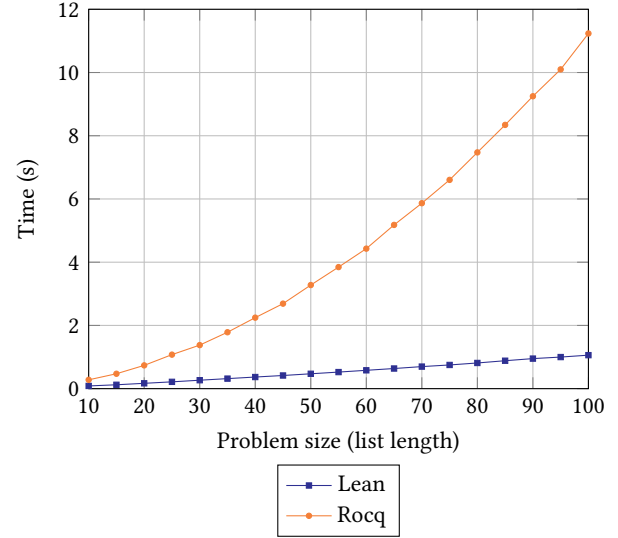

We use two sets of benchmarks to study the performance of the Iris Proof Mode in Lean against Iris-Rocq.\footnote{Iris-Lean revision \texttt{728a1714}, Iris-Rocq revision \texttt{a5375188}.} The measurements were collected on an AMD Ryzen AI 7 350, using Lean 4.32 without asynchronous elaboration and Rocq 9.2.0.

\paragraph{Framing}
First, to evaluate the performance of the core tactics and the custom typeclass synthesizer (\cref{sec:metaprogramming}), we create proof goals of the form $P~0 \sep \dots \sep P~n \vdash P~0 \sep \dots \sep P~n$ with varying $n$. Upon introducing and destructing the hypotheses using \leaninline|iintro|, they are canceled against the conclusion using \leaninline|iframe|. The same tests are also done with the conclusion consisting of the propositions in reverse order (\ie with entailments of the form $P~0 \sep \dots \sep P~n \vdash P~n \sep \dots \sep P~0$). The results of the benchmark are shown in \Cref{fig:FrameBenchmark}.
These two benchmarks test the two extremes of \leaninline|iframe|.
For the forward order, \leaninline|iframe| takes quadratic time since \leaninline|iframe| traverses the hypotheses from last introduced to first introduced (\ie from $P~n$ to $P~0$) and thus it needs to inspect all hypotheses until finding the one that matches the leftmost conjunct of the goal.
For the reverse order, the first hypothesis that \leaninline|iframe| inspects matches the predicate it searches for, avoiding a traversal of the whole context and leading to linear performance.
What is striking is the performance comparison with Iris-Rocq: even in the case where constant time is required for \rocqinline|iFrame|\footnote{The Iris-Rocq equivalent of \lstinline[basicstyle=\footnotesize\ttfamily]|iframe|.} to find a hypothesis, the performance in Rocq is still quadratic rather than linear, and thus the gap widens with $n$. For $n = 10$, it takes 4\,ms in Lean, as opposed to 7\,ms in Rocq; for $n = 120$, there is about a 27$\times$ difference (0.046\,s in Lean vs 1.256\,s in Rocq). A similar picture holds for the worst case, where Rocq exhibits cubic complexity against Lean's quadratic, so a 2.7$\times$ gap at $n = 10$ grows to nearly 9.0$\times$ difference at $n = 50$.

\paragraph{HeapLang verification}
Second, we evaluate the performance of program verification with a HeapLang program that consists of $n$ calls to a function and creates a list of size $n$. This tests the symbolic execution tactics of Iris-Lean and Iris-Rocq, in particular \leaninline|wp_pures|, \leaninline|wp_bind|, and \leaninline|iapply|. As shown in \Cref{fig:HeapLangBenchmark}, the performance of Iris-Lean is approximately linear, while it is superlinear in Iris-Rocq. Even for small sizes, the difference is significant, \eg around 3.3$\times$ for size 10 (86\,ms vs 279\,ms). The difference grows to around 10.6$\times$ at size 100 (1.06\,s vs 11.2\,s).

\paragraph{Performance differences}
It is hard to precisely pinpoint the causes for the performance differences between Iris-Rocq and Iris-Lean.
We speculate that a constant factor can be attributed to the fact that the context (the \leaninline|Hyps| object) is directly manipulated using meta-level Lean functions.
In contrast, Iris-Rocq represents the context as object-level lists that have to be reduced using reduction strategies like \rocqinline|cbv|, which accounts for a fifth of the runtime of \rocqinline|iFrame| in the benchmark.
For the asymptotic performance difference, we speculate that it is caused by better term sharing and caching of type checking in the Lean kernel:
while the algorithm of \leaninline|iframe| is linear for the reverse order in both Rocq and Lean, the proof term has quadratic size since the goal is repeated as a subterm at every step of the proof.
Our guess is that Lean can share the terms and type checking of these repeated occurrences of the goal, leading to linear performance, while Rocq repeatedly traverses every occurrence of the shared subterm, leading to quadratic performance.
While these benchmarks are synthetic, they suggest that Iris-Lean has the potential to deliver significant performance improvements for large-scale verification projects.

 \section{Conclusion}
\label{sec:conclusion}

\subsection{Related Work}

There have been a large number of mechanizations of various forms of separation logic inside proof assistants.
Several important early works were developed shortly after the initial development of separation logic itself.
\citet{DBLP:conf/csl/Weber04} gave the first machine-checked soundness proof in Isabelle of a separation logic for a simple language.
Although \citeauthor{DBLP:conf/csl/Weber04} was focused on verifying the meta-theory of the logic, subsequent work began to use mechanizations of separation logic to verify programs of interest directly in a theorem prover.
\citet{affeldt-topsy} applied a Rocq separation logic library to verify a memory allocator from an embedded operating system.
The VST project~\citep{appel2011vst} developed a separation logic library in Rocq for the Cminor language used in CompCert, thereby demonstrating that the approach of foundationally embedded separation logic could scale to complex features of a real programming language.
To do so, it incorporated a semantic model based on step indexing, as Iris later required for its advanced features.

Many separation logic frameworks, like Iris-Rocq and Iris-Lean, use a deep embedding to represent programs in the object language that are being verified.
However, other works have explored shallow embeddings that more tightly integrate with features of the meta-language.
Ynot~\citep{DBLP:conf/icfp/NanevskiMSGB08} is an axiomatic extension to Rocq based on Hoare Type Theory~\citep{DBLP:conf/icfp/NanevskiMB06}, which uses separation logic to reason about imperative monadic code.
CFML~\citep{chargueraud2011cfml} translated OCaml programs into a representation using characteristic formulas in Rocq.
These works used a shallow embedding for sequential programs.
The FCSL project~\citep{nanevski14fcsl} developed a shallow embedding in Rocq that incorporated concurrent separation logic features.
SteelCore~\citep{steelcore} and PulseCore~\citep{ebner2025pulse} support reasoning about shallowly embedded F* programs~\citep{swamy2016fstar}, and include support for challenging logical features like impredicative invariants, similar to those in Iris.

Productively using an embedded separation logic library requires a suite of tactics for carrying out proofs in the logic, as the Iris Proof Mode provides.
Several works have focused on the design of such tactic libraries.
\citet{DBLP:conf/tphol/McCreight09} developed interactive tactics for a separation logic for Cminor.
The Charge! framework~\citep{bengtson2012charge} included tactics for a higher-order separation logic.
Although these tactics provided an interactive experience, making it easier to apply the rules of the logic, they did not have the Iris Proof Mode representation of the separation logic context that could be manipulated using tactics analogous to the usual context in the proof assistant.
Bedrock~\citep{chlipala2011bedrock} instead focused on tactics for highly automating separation logic proofs.
Yolo~\citep{mikhalchuk2026yolo} implements automation for separation logics in Lean by separating proofs into two stages: a faster but unverified search phase followed by a slower proof reconstruction phase. An early version of Iris-Lean by König is used for a case study, where its CFML-style \leaninline|hProp| is mapped to Iris-Lean's BI.
We would be interested to see whether the two-stage approach can be extended to support our full Iris-Lean development.

IsarIris~\citep{sextl2022isariris} is a prototype implementation of Iris in the Isabelle/HOL proof assistant~\citep{nipkow2002isabelle}.
Like Iris-Lean, IsarIris attempts to closely adapt Iris-Rocq, using ML code generation to work around features that use dependent types such as \lstinline|inG|.
The port to Isabelle is incomplete, axiomatizing constructions like the America--Rutten fixed point theorem, although the authors claim that a proof is possible.
IsarIris does not adapt the MoSeL proof mode, adopting a fully custom backtracking search instead, though the authors conclude that this confers few advantages for automation.

Iris-Lean itself began with the initial proof mode foundations developed by \citet{koenig22}.
This initial development contained only a (partial) adaptation of the Iris Proof Mode and the generic bunched implications interface it uses.
The version of the project described in this paper represents several years of subsequent work by a large group of contributors.

\subsection{Future Work}
Our vision is for Iris-Lean to become a canonical piece of infrastructure for program verification in Lean.
To this end, we are exploring how Iris-Lean can best integrate Lean's new computer science library CSLib~\citep{barrett2026cslib} and with the \lstinline|Std.Do| program logic, which is being developed in Lean's standard library.
With the core library in hand, it is now possible to port specific program logics to Iris-Lean.

A long-term challenge remains in coordinating the future development of Iris-Lean and its relationship to Iris-Rocq.
While Iris-Lean has achieved parity with Iris-Rocq's current features, the latter continues to grow and evolve.
We hope to continue exploring ways in which we can take advantage of unique features of Lean to improve upon or simplify the implementation, while still trying to maintain some form of correspondence between the two developments.

\section*{Data Availability Statement}
The Lean mechanization accompanying this work is available on Zenodo~\citep{zenodo} and on GitHub at
\url{https://github.com/leanprover-community/iris-lean}.

\section*{Acknowledgments}
This work was supported in part by Villum Investigator grants (VIL73403 and VIL25804), Center for Basic Research in Program Verification (CPV), from the Villum Foundation.
This work was conducted in part while Markus de Medeiros was employed by Amazon Web Services and Math Inc. The views expressed herein are those of the authors and do not necessarily reflect those of Amazon Web Services, Math Inc., or their affiliates.
This work was supported in part by the \grantsponsor{NSF}{National Science Foundation}{}, grant no.~\grantnum{NSF}{2504143}.

\paragraph{Iris-Lean contributors}
This work was done as a combined effort of many contributors, and the project was largely driven by community effort.
We want to thank many people, who provided invaluable help with implementing Iris-Lean:
Alex Bai, 
Alex Keizer,
Alok Singh,
\mbox{@Garmelon},\\
\mbox{@GenericMonkey},
Haokun Li,
Joe Watt,
Klaus Kra{\ss}nitzer,
Lars K\"{o}nig,
L\'{e}o Stefanesco, 
\mbox{@Luke36},
Marcelo Fornet,
Puming Liu, 
Quang Dao, 
Remy Seassau,
Salkutsan Aleksey,
Sebastian Graf,
Seong-Heon Jung,
Shreyas Srinivas,
Sohail (Neel) Sarkar,
\mbox{@suhr},
Viet Anh Nguyen,
Yunsong Yang.

\end{document}